\documentstyle[11pt]{article}

\def\theequation{\arabic{section}.\arabic{equation}}

\renewcommand{\theequation}{\thesection.\arabic{equation}}

\global\arraycolsep=1pt
\input epsf
\input{tcilatex}
\begin{document}

\font\cmss=cmss10 \font\cmsss=cmss10 at 7pt \hfill \hfill IFUP-TH/02-34


\vspace{10pt}

\begin{center}
\vskip .5truecm{\Large {\bf \vspace{10pt}}}

{\Large {\bf SEARCH\ FOR\ FLOW INVARIANTS\ IN\ EVEN\ AND\ ODD\ DIMENSIONS}}

\bigskip \bigskip \vskip .5truecm

{\sl Damiano Anselmi}

{\it Dipartimento di Fisica E. Fermi, Universit\`{a} di Pisa, and INFN}
\end{center}

\vskip .5truecm

\begin{center}
{\sl Guido Festuccia}$^{\dagger }$

{\it Scuola Normale Superiore, Pisa, and INFN}

\vskip 2truecm

{\bf Abstract}
\end{center}

A flow invariant in quantum field theory is a quantity that does not depend
on the flow connecting the UV and IR conformal fixed points. We study the
flow invariance of the most general sum rule with correlators of the trace $%
\Theta $ of the stress tensor. In even (four and six) dimensions we recover
the results known from the gravitational embedding. We derive the sum rules
for the trace anomalies $a$ and $a^{\prime }$ in six dimensions. In three
dimensions, where the gravitational embedding is more difficult to use, we
find a non-trivial vanishing relation for the flow integrals of the three-
and four-point functions of $\Theta $. Within a class of sum rules
containing finitely many terms, we do not find a non-vanishing flow
invariant of type $a$ in odd dimensions. We comment on the implications of
our results.

\vskip 2truecm

{\small Pacs: 11.25.H; 11.10.Gh; 11.15.Bt; 11.40.Ex; 04.62.+v}

\vskip 4truecm

$^{\dagger }$ {\footnotesize Present address: Center for Theoretical
Physics, MIT, Cambridge, MA}\vfill\eject

\section{Introduction}

\setcounter{equation}{0}

The investigation of quantum field theory beyond the weakly coupled regime
is one of the difficult open problems of theoretical physics. The
renormalization group and anomalies provide powerful tools to study this
issue in several situations. Anomalies are quantities with special features,
which make them calculable to high orders in perturbation theory and
sometimes even exactly. The Adler--Bardeen theorem \cite{ABtheorem} is an
example of exact result (in the sense of the resummed perturbative
expansion) in quantum field theory. It has a variety of applications. For
example, it ensures that certain anomalies are one-loop exact. It also
justifies the 't Hooft anomaly matching conditions \cite{thooft}, which put
constraints on the strongly coupled limit of the theory.

The RG flow is intimately related to anomalies, the breaking of scale
invariance being encoded in the anomaly associated with the trace of the
stress tensor. In even dimensions, the trace anomaly in external gravity
defines the so-called {\sl central charges,} quantities which are
particularly useful to characterize the ultraviolet and infrared fixed
points of the RG\ flow. In supersymmetric theories, the two classes of
anomalies, axial and trace, are related to each other, and in several models
the Adler--Bardeen theorem can be used to compute the exact values of the
central charges in the interacting fixed points of the flows \cite{noi}.

A useful notion to understand the structure of the RG flow at a general
level is the notion of flow invariant. We can actually define two different
types of flow invariants, associated with the axial and trace anomalies,
respectively:

\noindent {\it i}) a flow invariant of the first type is a quantity $b$ that
does not depend on the energy scale, namely it is constant throughout the
RG\ flow, in particular it satisfies 
\[
b_{{\rm UV}}=b_{{\rm IR}}\ \text{;} 
\]
{\it ii})\ a flow invariant of the second type is a quantity $\Delta {\cal A}
$ that does not depend on the particular flow connecting the same pair of
UV\ and IR\ fixed points.

Examples of flow invariants of the first type are the 't Hooft anomalies
themselves. Consider a current $J_{\mu }$ associated with a classical
symmetry. The Adler-Bardeen theorem implies that if the current is conserved
at the quantum level (i.e. the internal anomaly vanishes), then all external
anomalies are one-loop exact. More explicitly, if the triangle $\left\langle
\partial J(x)\ J_{{\rm int}}(y)\ J_{{\rm int}}(z)\right\rangle $ vanishes
for all gauge currents $J_{{\rm int}}$, then the correlation function $%
\left\langle \partial J(x)\ J_{{\rm ext}}(y)\ J_{{\rm ext}}(z)\right\rangle $
is one-loop exact for all currents $J_{{\rm ext}}$ coupled to external
fields. This correlation function is uniquely determined by a constant $b$,
which therefore satisfies $b_{{\rm UV}}=b_{{\rm IR}}$.

Examples of flow invariants of the second type are provided by the central
charges. In particular, a flow invariant $\Delta {\cal A}$ is in general the
difference between the values of a central charge ${\cal A}$ at the fixed
points: 
\[
\Delta {\cal A}={\cal A}_{{\rm UV}}-{\cal A}_{{\rm IR}}. 
\]
In even dimensions well-known central charges are the quantities $c$ and $a$
defined by the trace anomaly in external gravity (details are given below).
The central charge $a$ satisfies the inequality $\Delta a\geq 0$ in higher
even-dimensional renormalizable unitary flows \cite{proc}. This property is
known as irreversibility of the RG flow. Instead, the sign of $\Delta c$ can
be negative. In two dimensions there is only one central charge ($c$ and $a$
coincide, in some sense) and does satisfy the irreversibility theorem \cite
{zamolo,cardy}. Finally, in odd dimensions the trace anomaly in external
gravity is identically zero at the critical points. A central charge $c$ can
be defined using the stress-tensor two-point function, but there is no clear
indication of the existence of a quantity like $a$.

In this paper, we report about the search for a central charge of type $a$
in odd dimensions, and the associated flow invariant. This research is meant
to start a program of systematic investigation of odd-dimensional quantum
field theory as an RG\ interpolation between conformal fixed points.

By central charge of type $a$ we mean a function ${\cal A}$ of the coupling
constant or, more generally, of the energy scale, with the following
properties:

\noindent 1) {\bf Flow invariance}, i.e. $\Delta {\cal A}={\cal A}_{{\rm UV}%
}-{\cal A}_{{\rm IR}}$ does not depend on the particular flow connecting
the\ fixed points.

\noindent 2) {\bf Stationariness} at the fixed points, i.e. 
\[
\left. \frac{\partial {\cal A}(\alpha )}{\partial \ln \alpha }\right|
_{\alpha \sim \alpha _{*}}\sim \beta _{\alpha },\qquad \text{or}\qquad
\left. \frac{\partial {\cal A}(\lambda )}{\partial \lambda }\right|
_{\lambda \sim \lambda _{*}}\sim \beta _{\lambda }. 
\]
Here $\alpha $ denotes a gauge coupling, while $\lambda $ denotes a coupling
of a Yukawa vertex, or $\varphi ^{4}$ in four dimensions, $\varphi^6$ in
three dimensions, etc. Formally, the formulas for gauge and non-gauge
couplings agree under the replacement $\ln \alpha \leftrightarrow \lambda $.
In particular, $\beta _{\alpha }={\rm d}\ln \alpha /{\rm d}\ln \mu $ and $%
\beta _{\lambda }={\rm d}\lambda /{\rm d}\ln \mu $. The star refers to the
critical values.

\noindent 3) {\bf Marginality}, i.e. ${\cal A}$ is constant on families of
continuously connected fixed points.

\noindent 4) {\bf Irreversibility}, i.e. $\Delta {\cal A}\geq 0$.

For the time being, we postpone the study of property 4). The first three
properties are sufficient to discriminate between central charges of type $c$
and central charges of type $a$, at least in four dimensions \cite{ccfis,noi}%
. We look for a quantity satisfying both 1), 2) and 3). More precisely, we
study a class of sum rules for $\Delta {\cal A}$ satistying 2) and check if
they satisfy also 1) and 3). The result is that in even dimensions we do not
find more flow invariants than the known one, $\Delta a$, while in odd
dimensions we do not find a genuine flow invariant of type $a$, but we do
find a nontrivial vanishing relation, that is to say a flow invariant of
type ($ii$) identically equal to zero.

In odd dimensions, the embedding in external gravity is not helpful to
define a quantity $a$, because the trace anomaly of the theory embedded in
external gravity identically vanishes at the fixed points. The reason is
that the dimension of the stress tensor in units of mass equals the
space-time dimension, but no scalar of odd dimension can be constructed with
the Riemann tensor and its covariant derivatives. On the other hand, there
is evidence, both in even and odd dimensions, that the properties of the
central charges $c$ and $a$ and the flow invariants $\Delta c$ and $\Delta a$
associated with them need more than the embedding in external gravity, to be
explained. For example, in even dimensions the subclass of conformal field
theories having $c=a$ is singled out by a special structure ${\cal G}_{d}$
in the trace anomaly \cite{proc,cea}. The peculiarity of the combination $%
{\cal G}_{d}$ does not appear to follow from the properties of the embedding
in external gravity. Second, the flows having $\Delta a=\Delta a^{\prime }$
are associated with a ``pondered'' extension of the Euler density \cite
{athm,proc}, whose existence, again, does not appear to follow from the mere
existence of a gravitational embedding. Third, in Gaussian massive models
there appeas to be a nontrivial relation between $a^{\prime }$ and $c$ \cite
{inv,proc}, both in even and in odd dimensions, but from the theoretical
point of view this relation remains to be fully explained. The embedding in
external gravity does not seem to be able to provide the missing
understandings.

These observations suggest that the explanation of the properties of central
charges and flow invariants of type ({\it ii}) must lie beyond the
gravitational embedding. Presumably, every property can be derived without
using the gravitational embedding. In this paper, we show that the quantity $%
a$ can be singled out, in even dimensions, using only flow invariance. We
exploit this fact to seek for a similar quantity in odd dimensions.

Our empirical approach, based on the search for flow invariants in specific
toy models without using the embedding in external gravity, is not
sufficient to rigorously prove the existence of flow invariants.
Nevertheless, this is the most economical method we have at the moment to
guide the search for flow invariants in odd dimensions. Note that the
vanishing relations and sum rules of this paper and ref. \cite{234} are
universal, that is to say they hold in every quantum field theory
(renormalizable and non-renormalizable, unitary and non-unitary)
interpolating between well-defined UV and IR fixed points.

It is impossible to mention here all of the works that can be found in the
literature about correlators of composite operators, in particular the
stress tensor and its trace, with or without the embedding in external
fields. We just point out that the idea of working without the gravitational
embedding is not new in itself. What is new is the context in which we use
it, in odd dimensions, and the purpose. Curiously, however, in the end we
discover the relation (\ref{van}) in three dimensions, whose features
suggest that it does follow from the gravitational embedding. It would be
interesting to rederive (\ref{van}) from the gravitational embedding. This
is more difficult than in even dimensions and we exect that the techniques
of \cite{234} need deep improvements.

\bigskip

We now illustrate the plan of the research and the method.

We assume that the stress tensor is multiplicatively renormalizable, i.e.
that there is no improvement term. Then the stress tensor is finite. In the
presence of improvements, the treatment can be generalized using the minimum
principle of section 7 of \cite{inv}. Moreover, at some point in the
argument below we specialize, for pedagogical reasons, to classically
conformal theories, where the trace of the stress tensor is an
``evanescent'' operator (because it is classically zero) and its
renormalized expression has the form 
\begin{equation}
\Theta =\beta _{j}{\rm O}_{j}  \label{betao}
\end{equation}
for certain operators O$_{j}$. For example, O$=-F^{2}/4$ in Yang-Mills
theory and O$=\varphi ^{4}/4!$ in the $\varphi ^{4}$-theory in four
dimensions. The formulas below are written for a gauge coupling, for
simplicity. The conclusions that we derive here hold also in the more
general cases.

To have 2)\ guaranteed, we can assume that the function ${\cal A}(\alpha
(\mu ))$, or ${\cal A}(t)$, where $t=\ln \mu $, is such that, roughly 
\[
\frac{{\rm d}{\cal A}}{{\rm d}t}\sim \int \left\langle \Theta \Theta
\right\rangle +\int \left\langle \Theta \Theta \Theta \right\rangle +\cdots
,\qquad \frac{{\rm d}{\cal A}}{{\rm d}t}\sim \beta ^{2},\qquad \frac{%
\partial {\cal A}}{\partial \ln \alpha }\sim \beta _{\alpha }. 
\]
This assures stationariness at the critical points. Below we make this
argument rigorous, helping ourselves with the embedding in external gravity,
eventually specializing to a conformally flat external metric. First we give
some details on the notation and the scheme conventions.

We consider the correlators

\begin{equation}
\left\langle \Theta (x_{1})~\cdots ~\Theta (x_{i})~\Theta (0)\right\rangle .
\label{corr}
\end{equation}
In general, these correlators need subtractions at the coincident points
(see \cite{hathrell} for a detailed treatment of this issue in perturbation
theory). In our case, we can deal with these subtractions as follows. We
distinguish the subtractions of ``overall'' contact terms (i.e. the local
terms where all of the points coincide) from the subtractions of semilocal
terms. The subtractions of semilocal terms cancel out if we replace (\ref
{corr}) with the functional derivatives $-\Gamma _{x_{1}\cdots x_{i}0}$ of
the induced action $\Gamma [\phi ]$ with respect to the conformal factor $%
\phi $ at the points $x_{1},\cdots ,x_{i},0$. The reason is that $\Gamma
[\phi ]$ is finite (after the subtraction of appropriate counterterms
constructed with the metric) and the functional derivatives of $\Gamma [\phi
]$ are therefore finite \cite{hathrell}. The overall local terms are
associated with the counterterms constructed with the metric. It was shown
in \cite{234} that if we calculate (\ref{corr}) in the framework of the
dimensional-regularization technique and the minimal subtraction scheme,
with a conformally flat external metric, the overall local terms vanish.
Then the finite (anomalous) terms that have the appropriate structure and
dimensionality to mix with (\ref{corr}) are collected in the ultraviolet
limit $\Gamma _{{\rm UV}}$ of $\Gamma [\phi ]$. In even dimensions, $\Gamma
_{{\rm UV}}$ depends only on $a$ and $a^{\prime }$. Expressions of $\Gamma _{%
{\rm UV}}[\phi ]$ in four and six dimensions are (\ref{boh}), (\ref{boha}), (%
\ref{6corr}) and (\ref{6cin}). In odd dimensions, instead, $\Gamma _{{\rm UV}%
}[\phi ]$ is zero, because the trace anomaly vanishes at the fixed points.

In conclusion, with our conventions it is appropriate to replace (\ref{corr}%
) with the functional derivatives $-\Gamma _{x_{1}\cdots x_{i}0}^{\prime }$
of the difference $\Gamma ^{\prime }=\Gamma -\Gamma _{{\rm UV}}$ \cite{234}: 
\begin{eqnarray}
\Gamma _{x}^{\prime } &=&-\langle \overline{\Theta }(x)\rangle ,\qquad
\qquad \Gamma _{x_{1}x_{2}}^{\prime }[\phi ]=-\langle \overline{\Theta }%
(x_{1})\,\overline{\Theta }(x_{2})\rangle -\langle \overline{\Theta }%
_{x_{2}}(x_{1})\rangle ,  \label{dera} \\
\Gamma _{x_{1}x_{2}x_{3}}^{\prime }[\phi ] &=&-\langle \overline{\Theta }%
(x_{1})\,\overline{\Theta }(x_{2})\,\overline{\Theta }(x_{3})\rangle
-\langle \overline{\Theta }_{x_{3}}(x_{1})\,\overline{\Theta }(x_{2})\rangle
-\langle \overline{\Theta }(x_{1})\,\overline{\Theta }_{x_{3}}(x_{2})\rangle
\nonumber \\
&&~~~~~~~~~~~~~~~~-\langle \overline{\Theta }_{x_{2}}(x_{1})\,\overline{%
\Theta }(x_{3})\rangle -\langle \overline{\Theta }_{x_{2}x_{3}}(x_{1})%
\rangle ,\qquad \qquad {\rm etc}.  \nonumber
\end{eqnarray}
Here $\overline{\Theta }_{x_{1}\cdots x_{i}}(x)\,$ denote the functional
derivatives of the operator $\overline{\Theta }(x)=\sqrt{g}\Theta =-\delta
S/\delta \phi (x)$ with respect to $\phi (x_{1})\cdots \phi (x_{i})$, and $S$
is the action embedded in external gravity. The external metric is set to be
flat after differentiation.

\bigskip

We postulate sum rules of the form 
\begin{eqnarray}
\Delta {\cal A} &=&-\int {\rm d}^{d}x\,P_{1}(x)\,\Gamma _{x0}^{\prime }-\int 
{\rm d}^{d}x\,{\rm d}^{d}y\,P_{2}(x,y)\,\Gamma _{xy0}^{\prime }-\int {\rm d}%
^{d}x\,{\rm d}^{d}y\,{\rm d}^{d}z\,P_{3}(x,y,z)\Gamma _{xyz0}^{\prime
}+\cdots  \nonumber \\
&=&-\sum_{i=1}^{\infty }\int P_{i}(x_{1},\cdots ,x_{i})\,\Gamma
_{x_{1}\cdots x_{i}0}^{\prime }~\prod_{k=1}^{i}{\rm d}^{d}x_{k},
\label{summa}
\end{eqnarray}
where the $P_{i}(x_{1},\cdots ,x_{i})$s are Lorentz-invariant (SO($d$%
)-invariant, in the Euclidean framework) homogeneous functions of degree $d$
in the coordinates. In even dimensons, the $P_{i}$s are polynomials and, due
to certain vanishing relations \cite{234}, the sum (\ref{summa}) is finite.
The $P_{i}$s are model-independent, and such that the integrals of (\ref
{summa}) are convergent.

We insert identities 
\[
1=\int_{0}^{\infty }{\rm d}\zeta ^{2}~\delta \left( \zeta ^{2}-F_{i}\left(
x_{1},\cdots ,x_{i}\right) \right) , 
\]
where the $F_{i}$s are positive homogeneous functions of degree two in the $%
x_{1},\cdots ,x_{i}$. For example, we can take 
\[
F_{i}\left( x_{1},\cdots ,x_{i}\right) =\sum_{k=1}^{i}|x_{k}|^{2}. 
\]
We then perform the rescaling $x_{k}\rightarrow \zeta x_{k}$, and use the
Callan-Symansik equations and the finiteness of $\Theta $. We obtain 
\begin{equation}
\Delta {\cal A}=-2\int_{0}^{\infty }\frac{{\rm d}\zeta }{\zeta }%
~\sum_{i=1}^{\infty }\int \prod_{k=1}^{i}{\rm d}^{d}x_{k}~P_{i}(x_{1},\cdots
,x_{i})\delta \left( 1-F_{i}\left( x_{1},\cdots ,x_{i}\right) \right)
\,\Gamma _{x_{1}\cdots x_{i}0}^{\prime }(\alpha (\zeta \mu )).
\label{bibidi}
\end{equation}

With $t=\ln \zeta $ and changing notation from $\alpha (\zeta \mu )$ to $%
\alpha (t)$, we can define the function ${\cal A}(t)$ such that 
\begin{eqnarray}
\dot{{\cal A}}(t) &=&2\sum_{i=1}^{\infty }\int P_{i}(x_{1},\cdots
,x_{i})\delta \left( 1-F_{i}\left( x_{1},\cdots ,x_{i}\right) \right)
\,\Gamma _{x_{1}\cdots x_{i}0}^{\prime }(\alpha (t))\prod_{k=1}^{i}{\rm d}%
^{d}x_{k},  \label{bibidibo} \\
\Delta {\cal A} &=&-\int_{-\infty }^{+\infty }{\rm d}t\ \dot{{\cal A}}(t)=%
{\cal A}(-\infty )-{\cal A}(\infty ),\qquad {\cal A}(t)={\cal A}(-\infty
)+\int_{-\infty }^{t}{\rm d}t^{\prime }\,\dot{{\cal A}}(t^{\prime }). 
\nonumber
\end{eqnarray}
If the theory is classically conformal, $\mu $ is the unique scale at the
quantum level and ${\cal A}(t)$ depends only on $t$. Otherwise, the function 
${\cal A}(t)$ depends also on the ratios between $\mu $ and the other
dimensioned parameters, typically the masses. In even dimensions, ${\cal A}%
(\pm \infty )$ are both zero (vanishing relations) or the UV and IR values
of the central charge $a$. In odd dimensions only the difference ${\cal A}%
(t)-{\cal A}(-\infty )$ is defined.

Let us focus on classically conformal theories. Here, the terms of (\ref
{dera}) containing the derivatives $\overline{\Theta }_{x_{1}\cdots x_{i}}$
of $\overline{\Theta }$ are evanescent and can be neglected in the sum rules 
\cite{234}, if the integrals (\ref{summa}) converge, which we have assumed.
In this particular case, $\Gamma _{x_{1}\cdots x_{i}0}^{\prime }$ can be
replaced with minus (\ref{corr}). Then, it is simple to prove that the
function ${\cal A}(t)$ is stationary at criticality. Indeed, from the
considerations made so far, equation (\ref{betao}) tells us that 
\begin{equation}
\left\langle \Theta (x_{1})~\cdots ~\Theta (x_{i})~\Theta (0)\right\rangle
=\beta ^{i+1}(\alpha )\ \left\langle {\rm O}(x_{1})~\cdots ~{\rm O}(x_{i})~%
{\rm O}(0)\right\rangle .  \label{blabla}
\end{equation}
We recall that in general this correlator should contain additional local
terms from the conformal anomaly, but these terms have been shifted to the
definition of the $\Theta $ -correlators, according to our notation and
conventions. The O-correlators are regular throughout the RG flow. We
conclude from (\ref{bibidibo}) that 
\begin{equation}
\frac{\partial {\cal A}(\alpha )}{\partial \ln \alpha }=2\sum_{i=1}^{\infty
}\beta ^{i}(\alpha )\int \prod_{k=1}^{i}{\rm d}^{d}x_{k}\ P_{i}(x_{1},\cdots
,x_{i})\delta \left( 1-F_{i}\left( x_{1},\cdots ,x_{i}\right) \right)
\,\left\langle {\rm O}(x_{1})~\cdots ~{\rm O}(x_{i})~{\rm O}(0)\right\rangle
.  \label{bubudu}
\end{equation}
This formula contains one factor of $\beta $ less than (\ref{bibidibo}),
since $\dot{{\cal A}}(t)=-\beta \ \partial {\cal A}(\alpha )/\partial \ln
\alpha $. The surviving factors of $\beta $ are however sufficient to prove
that (\ref{bubudu}) vanishes at both critical points.

\medskip

In odd dimensions the functions $P_{i}$s, which have dimension $d$, are not
polynomial. A priori, the sum rule (\ref{summa}) could contain infinitely
many terms. Some nontrivial restrictions are provided by the requirement
that the integrals be convergent. We will also see in a concrete
one-dimensional example that our approach does not allow us to study sum
rules with infinitely many terms. We are forced to put some restrictions on
the $P_{i}$s. For example that they do not contain negative powers of the
coordinates and that they are sufficiently simple. Then, the surviving set
of $P_{i}$s is finite. The set we choose does not single out a flow
invariant in odd dimensions, but exhibits the existence of a non-trivial
vanishing relation among flow integrals of the three- and four-point
correlators.

\medskip

The paper is organized as follows. In section 2 we describe the procedure in
even dimensions, four and six. We show that it is possible to ignore the
existence of a coupling to external gravity and fix (\ref{summa}) so that it
is flow invariant. We show that the unique non-trivial flow invariant is the
one associated with the central charge $a$. In section 3 we study the
problem in odd dimensions, three and one. We show that the simplest class of 
$P_{i}$s does not produce any flow invariant, but they give the non-trivial
vanishing relation (\ref{van}), which might be implied by the coupling to
external gravity. In the conclusions we comment on the possibilities that
remain open after our analysis.

\section{Search for flow invariants in even dimensions}

Before describing our calculations, let us recall how the embedding in
external gravity defines the central charges $c$, $a$ and $a^{\prime }$ in
even dimensions $d=2n$.

The trace anomaly at criticality in even dimensions contains three types of
terms constructed with the curvature tensors and their covariant
derivatives: {\it i}) terms ${\cal W}_{i}$, $i=0,1,\ldots ,I$, such that $%
\sqrt{g}{\cal W}_{i}$ are conformally invariant; {\it ii}) the Euler density 
\[
{\rm G}_{d}=(-1)^{n}\varepsilon _{\mu _{1}\nu _{1}\cdots \mu _{n}\nu
_{n}}\varepsilon ^{\alpha _{1}\beta _{1}\cdots \alpha _{n}\beta
_{n}}\prod_{i=1}^{n}R_{\alpha _{i}\beta _{i}}^{\mu _{i}\nu _{i}}~; 
\]
{\it iii}) covariant total derivatives ${\cal D}_{j}$, $j=0,1,\ldots ,J$,
having the form $\nabla _{\alpha }J^{\alpha }$, $J^{\alpha }$ denoting a
covariant current.

The coefficients multiplying these terms in the trace anomaly are denoted
with $c_{d}^{i}$, $a_{d}$ and $a_{d}^{j~\prime }$, respectively. We write $%
c_{d}=c_{d}^{0}$ and $a_{d}^{\prime }=a_{d}^{0~\prime }$. We have 
\begin{equation}
{\Theta _{d{\rm =}2n}^{*}={\frac{{n}!}{(4\pi )^{n}\,(d+1)!}}\left[ {\frac{%
c_{d}\left( d-2\right) }{4(d-3)}}{\cal W}_{0}+\sum_{i=1}^{I}c_{d}^{i}{\cal W}%
_{i}-{\frac{2^{1-n}}{\,d}}\left( a_{d}{\rm G}_{d}+\sum_{j=0}^{J}a_{d}^{j~%
\prime }{\cal D}_{j}\right) \right] .}  \label{trace}
\end{equation}
Here $c_{2}$ is $a_{2}$. ${\cal W}_{0}$ is the unique term of the form $%
W\Box ^{n-2}W+\cdots $ such that $\sqrt{g}{\cal W}_{0}$ is conformally
invariant, where the dots denote cubic terms in the curvature tensors. I
have separated ${\cal W}_{0}$ from the other terms of the type ${\cal W}_{i}$%
, because its coefficient $c_{d}$ is also the coefficient of the
stress-tensor two-point function: 
\begin{eqnarray}
\langle T_{\mu \nu }(x)\,T_{\rho \sigma }(0)\rangle &=&c_{d}{\frac{\Gamma
\left( n+1\right) \Gamma \left( n\right) }{8\pi ^{d}(d+1)d(d-1)(d-2)^{2}}%
\prod }_{\mu \nu ,\rho \sigma }^{(2)}\left( {\frac{1}{|x|^{2d-4}}}\right) ,
\label{cc} \\
{\prod }_{\mu \nu ,\rho \sigma }^{(2)} &=&{\frac{1}{2}}(\pi _{\mu \rho }\pi
_{\nu \sigma }+\pi _{\mu \sigma }\pi _{\nu \rho })-{\frac{1}{d-1}}\pi _{\mu
\nu }\pi _{\rho \sigma },~~~~~~~~~\pi _{\mu \nu }=\partial _{\mu }\partial
_{\nu }-\Box \delta _{\mu \nu },  \nonumber
\end{eqnarray}
and is normalized so that for free fields ($n_{s}$ real scalars, $n_{f}$
Dirac fermions and, in even dimensions, $n_{v}$ ($n-1$)-forms) it reads 
\begin{equation}
c_{d}=n_{s}+2^{[n]-1}(d-1)n_{f}+\frac{d!}{2\left[ \left( n-1\right) !\right]
^{2}}n_{v}.  \label{freec}
\end{equation}
Definition (\ref{cc}) and formula (\ref{freec}) hold also if the spacetime
dimension $d$ is odd (but greater than one), if $n_{\nu }$ is set to zero
and $[n]$ denotes the integral part of $n$. The covariant total-derivative
term 
\[
{\cal D}_{0}=-{\frac{2^{n}d}{2(d-1)}}\Box ^{n-1}R, 
\]
is also peculiar among the ${\cal D}_{j}$s, since it is the unique t${\cal D}
$ linear in the curvature tensors. We choose a basis such that all the $%
{\cal D}_{j}$, $j>0$, are at least quadratic in the curvature tensors. Then,
on conformally flat metrics, ${\cal D}_{0}$ contains the unique term of
formula (\ref{trace}) that is linear in the conformal factor $\phi $.

The quantities $\Delta a_{d}^{j~\prime }$ are not flow invariants of type $%
ii $), because the $a_{d}^{j~\prime }$s are ill-defined at criticality.
Arbitrary, RG-invariant constants $\delta a_{d}^{j\,\prime }$ can be added
to these quantities. These ambiguities are due to the arbitrary additions of
finite local terms ${\bf D}_{j}$ to the action $\Gamma $: 
\begin{equation}
\Gamma \rightarrow \Gamma -{\frac{2^{1-n}n!}{(4\pi )^{n}(d+1)!d}}\delta
a_{d}^{j\,\prime }\int {\rm d}^{2n}x\,\sqrt{g}{\bf D}_{j}.  \label{ambi}
\end{equation}
The ${\cal D}_{j}$s are just the $\phi $-derivatives of the ${\bf D}_{j}$s: 
\[
\sqrt{g}{\cal D}_{j}(x)=-{\frac{\delta }{\delta \phi (x)}}\int {\rm d}%
^{2n}x\,\sqrt{g}{\bf D}_{j}. 
\]
The ambiguities disappear in the differences $\Delta a_{d}^{j\,\prime }$,
which are, however, no longer flow invariant.

The central charges can be normalized universally, as shown in ref. \cite
{proc}. First, the central charge $c_{d}$ is set to be one for the free real
scalar field. Secondly, the relative normalization of $c_{d}$, the other $%
c_{d}^{i}$s and $a_{d}$ is determined by the trace-anomaly structure ${\cal G%
}_{d}$ that defines the ``$c=a$'' theories of ref. \cite{cea}. Third, the
relative normalization of $a_{d}$ and the $a_{d}^{j\,\prime }$s is
determined by the pondered Euler density of ref. \cite{at6d}.

\subsection{Four dimensions}

In four dimensions we have 
\[
\Theta _{d{\rm =4}}^{*}=\frac{1}{120}{\frac{1}{(4\pi )^{2}}}\left[ c\,W^{2}-{%
\frac{a}{4}}\,{\rm G}_{4}+{\frac{2}{3}}\,a^{\prime }\,\Box R\right] 
\]
and the sum rules \cite{234}

\begin{eqnarray}
\Delta a &=&-{\frac{5\pi ^{2}}{2}}\int {\rm d}^{4}x\,|x|^{4}\,\Gamma
_{x0}^{\prime }-{\frac{5\pi ^{2}}{2}}\int {\rm d}^{4}x\,{\rm d}%
^{4}y\,x^{2}\,y^{2}\,\Gamma _{xy0}^{\prime }  \nonumber \\
&=&-{\frac{5\pi ^{2}}{2}}\int {\rm d}^{4}x\,|x|^{4}\,\Gamma _{x0}^{\prime }-{%
\frac{5\pi ^{2}}{2}}\int {\rm d}^{4}x\,{\rm d}^{4}y\,{\rm d}^{4}z\,\left(
x\cdot y\right) \left( x\cdot z\right) \Gamma _{xyz0}^{\prime }.
\label{equa}
\end{eqnarray}

These sum rules have been derived using the coupling to external gravity. To
prepare the search for flow invariants in odd dimensions, where the coupling
to external gravity is more difficult to use, we derive the sum rules in an
alternative way. We postulate sum rules of the form (\ref{summa}) and
determine the polynomials $P_{i}$s so that $\Delta {\cal A}$ is a flow
invariant.

Let us fix restrictions on the polynomials $P_{i}(x_{1},\ldots ,x_{i})$.
First, we know on dimensional grounds that the polynomials have degree four.
They have of course to be Lorentz invariant. It was shown in \cite{234} that
there exist equivalence relations among polynomials. In particular, there
exist one irreducible monomial for $i=1$, which is $|x_{1}|^{4}$, one for $%
i=2$, which is for example $x_{1}^{2}x_{2}^{2}$, one for $i=3$, for example $%
\left( x_{1}\cdot x_{2}\right) \left( x_{1}\cdot x_{3}\right) $, and none
for $i>3$. The most general sum rule we have to inspect reads therefore 
\begin{equation}
\Delta {\cal A}=-{\frac{5\pi ^{2}}{2}}\left[ \lambda _{1}\int {\rm d}%
^{4}x\,|x|^{4}\,\Gamma _{x0}^{\prime }+\lambda _{2}\int {\rm d}^{4}x\,{\rm d}%
^{4}y\,x^{2}\,y^{2}\,\Gamma _{xy0}^{\prime }+\lambda _{3}\int {\rm d}^{4}x\,%
{\rm d}^{4}y\,{\rm d}^{4}z\,\left( x\cdot y\right) \left( x\cdot z\right)
\Gamma _{xyz0}^{\prime }\right] .  \label{sumj}
\end{equation}
Of the three unknown constants $\lambda _{1}$, $\lambda _{2}$, $\lambda _{3}$%
, one is an overall normalization.

The three terms of (\ref{sumj}) are in one-to-one correspondence with the
terms of the most general local expression for the effective action $\Gamma
[\phi ]$ at criticality. We restrict to conformally flat metrics $g_{\mu \nu
}=\delta _{\mu \nu }{\rm e}^{2\phi }$ and study only the dependence of $%
\Gamma $ on the conformal factor $\phi $. Neglecting total derivatives,
power counting gives 
\begin{equation}
\Gamma ^{*}[\phi ]=\int {\rm d}^{4}x\{b_{1}(\Box \phi )^{2}+b_{2}(\Box \phi
)(\partial _{\mu }\phi )^{2}+b_{3}(\partial _{\mu }\phi )^{4}\}  \label{boh}
\end{equation}
If we knew that the action $\Gamma ^{*}[\phi ]$ was inherited from the
coupling to external gravity, we would have only two independent terms,
corresponding to $a$ and $a^{\prime }$, instead of three \cite{athm}: 
\begin{equation}
\Gamma ^{*}[\phi ]=\frac{1}{60}\frac{1}{(4\pi )^{2}}\int {\rm d}^{4}x\left\{
a_{*}(\Box \phi )^{2}-(a_{*}-a_{*}^{\prime })\left[ \Box \phi +(\partial
_{\mu }\phi )^{2}\right] ^{2}\right\} .  \label{boha}
\end{equation}
We ignore this restriction for the moment and derive it from flow invariance.

From (\ref{boh}) we can derive a set of sum rules for the $\Delta b_{i}$,
following the procedure of \cite{234}. The results are 
\begin{eqnarray*}
&&\int x^{4}\;\Gamma _{x,0}^{\prime }\;{\rm d}^{4}x=-384\Delta b_{1}, \\
&&\int (x^{2}y^{2})\;\Gamma _{x,y,0}^{\prime }\;{\rm d}^{4}x\;{\rm d}%
^{4}y=192\Delta b_{2}, \\
&&\int (x\cdot y)(x\cdot z)\;\Gamma _{x,y,z,0}^{\prime }\;{\rm d}^{4}x\;{\rm %
d}^{4}y\;{\rm d}^{4}z=384\Delta b_{3}.
\end{eqnarray*}

We inspect the sum rules in the case of the higher-derivative scalar field,
with action

\begin{equation}
S=\frac{1}{2}\int {\rm d}^{2n}x\sqrt{g}[\varphi \Delta _{4}\varphi +\beta
m^{2}(\partial _{\mu }\varphi )(\partial _{\nu }\varphi )g^{\mu \nu
}+m^{4}\varphi ^{2}+\eta Rm^{2}\varphi ^{2}],  \label{azionenonunit}
\end{equation}
here written in generic spacetime dimension $d=2n$ for later convenience.
The field $\varphi $ has dimension $\left( d-4\right) /2$ in units of mass.
The fourth-order differential operator $\Delta _{4}$ (see for example \cite
{johanna}) 
\begin{eqnarray}
\Delta _{4} &=&\nabla ^{2}\nabla ^{2}+\nabla _{\mu }\left[ {\frac{4}{d-2}}%
R^{\mu \nu }-{\frac{d^{2}-4d+8}{2(d-1)(d-2)}}g^{\mu \nu }R\right] \nabla
_{\nu }-{\frac{d-4}{4(d-1)}}\nabla ^{2}R  \nonumber \\
&&-{\frac{d-4}{(d-2)^{2}}}R_{\mu \nu }R^{\mu \nu }+{\frac{%
(d-4)(d^{3}-4d^{2}+16d-16)}{16(d-1)^{2}(d-2)^{2}}}R^{2}  \label{due}
\end{eqnarray}
transfoms as 
\[
\Delta _{4}\rightarrow {\rm e}^{-(n+2)\phi }\Delta _{4}{\rm e}^{(n-2)\phi } 
\]
under a Weyl rescaling $g_{\mu \nu }\rightarrow {\rm e}^{2\phi }g_{\mu \nu }$%
.

We write $\beta =r+1/r$, where $r^{2}$ is the ratio between the two poles of
the propagator. Indeed, (\ref{azionenonunit}) reads in flat space 
\[
S=\frac{1}{2}\int {\rm d}^{2n}x\ \varphi \left( -\Box +rm^{2}\right) \left(
-\Box +m^{2}/r\right) \varphi , 
\]
which means that the theory propagates two fields (one of which is a ghost)
with masses $rm^{2}$ and $m^{2}/r$. The results can be taken from \cite{impr}%
: 
\begin{eqnarray*}
-{\frac{5\pi ^{2}}{2}}\int {\rm d}^{4}x\,|x|^{4}\,\Gamma _{x0}^{\prime }
&=&u(r)=960\pi ^{2}\Delta b_{1}, \\
-{\frac{5\pi ^{2}}{2}}\int {\rm d}^{4}x\,{\rm d}^{4}y\,x^{2}\,y^{2}\,\Gamma
_{xy0}^{\prime } &=&-\frac{28}{3}-u(r)=-480\pi ^{2}\Delta b_{2}, \\
-{\frac{5\pi ^{2}}{2}}\int {\rm d}^{4}x\,{\rm d}^{4}y\,{\rm d}^{4}z\,\left(
x\cdot y\right) \left( x\cdot z\right) \Gamma _{xyz0}^{\prime } &=&-\frac{28%
}{3}-u(r)=-960\pi ^{2}\Delta b_{3},
\end{eqnarray*}
where 
\[
u(r)=3{\frac{1+17r^{2}-17r^{4}-r^{6}+10(1+r^{2}+r^{4}+r^{6})\ln r}{%
(r^{2}-1)^{3}}.} 
\]
We have therefore 
\[
\Delta {\cal A}=\left( \lambda _{1}+\lambda _{2}+\lambda _{3}\right) u(r)-%
\frac{28}{3}\left( \lambda _{2}+\lambda _{3}\right) . 
\]
Demanding flow invariance, that is to say independence of $r$, we find 
\[
\lambda _{1}+\lambda _{2}+\lambda _{3}=0, 
\]
and therefore 
\begin{equation}
\Delta {\cal A}=-\frac{28}{3}\lambda _{1}.  \label{auto}
\end{equation}
Now if $\lambda _{1}\neq 0$, this is precisely the first formula of (\ref
{equa}) for $\Delta a$ (for $\lambda _{1}=1$). If, instead, $\lambda _{1}=0$%
, then $\lambda _{3}=-\lambda _{2}$ and we have a null flow invariant, that
is to say a non-trivial vanishing relation ($\Delta b_{2}=2\Delta b_{3}$)
which does not follow from the kinematic sum rules (see discussion about
this point in \cite{234}). This relation is the difference between the two
expressions (\ref{equa}) for $\Delta a$ and explains the restriction from (%
\ref{boh}) to (\ref{boha}).

We conclude that there are two flow invariants. Having explored only one
explicit model, we can always arrange the invariants in one non-vanishing
invariant, plus a remainder of vanishing invariants. Only exploring a larger
class of models, for example the higher-derivative fermion, we can have true
evidence that the non-vanishing invariant is unique. We will not do this
here and proceed in other directions.

\subsection{Six dimensions}

We now repeat the procedure in six dimensions and confirm the conlusions
derived in the previous section. In six dimensions the form of the anomaly
at criticality $\Theta ^{*}$ is \cite{at6d} 
\begin{equation}
{\Theta _{6}^{*}={\frac{{1}}{840(4\pi )^{3}\,}}\left[ {\frac{c}{3}}{\cal W}%
_{0}+c_{1}{\cal W}_{1}+c_{2}{\cal W}_{2}-{\frac{1}{\,24}}\left( a{\rm G}%
_{6}+a^{\prime }{\cal D}_{0}+\sum_{j=1}^{5}a_{j}^{~\prime }{\cal D}%
_{j}\right) \right] }.  \label{anom6}
\end{equation}
The conformally invariant terms are 
\begin{eqnarray*}
{\cal W}_{0} &=&W_{\mu \nu \rho \sigma }\left( \Box \delta _{\alpha }^{\mu
}+4R_{\alpha }^{\mu }-\frac{6}{5}R\delta _{\alpha }^{\mu }\right) W^{\alpha
\nu \rho \sigma }+\nabla _{\mu }J^{\mu }, \\
{\cal W}_{1} &=&W_{\mu \nu \rho \sigma }W^{\mu \nu \alpha \beta }W_{\alpha
\beta }^{\rho \sigma },\qquad \qquad {\cal W}_{2}=W_{\mu \nu \rho \sigma
}W^{\mu \alpha \rho \beta }W_{\alpha \space \beta }^{\nu \space \sigma },
\end{eqnarray*}
where the precise expression of $J^{\mu }$ can be found in \cite
{johanna,bastia}. The total-derivative terms ${\cal D}_{j}$ have been
classified by Bastianelli {\it et al.} in \cite{bastia}, where it was shown
that there are six such terms. For our purposes, we are interested only in
the three ${\cal D}_{j}$s that do not vanish on a conformally flat metric.
We choose a basis made by ${\cal D}_{0}$ and 
\[
{\cal D}_{1}=\frac{234}{25}\Box ^{2}R,\qquad \qquad {\cal D}_{2}=-12\Box
R_{\mu \nu }^{2}-\frac{24}{5}\Box R^{2}-\frac{48}{5}\nabla _{\mu }\left(
R^{\mu \nu }\nabla _{\nu }R\right) . 
\]
We recall that for conformally flat metric $W=0$ and the only unambiguous
term is $aG_{6}$. It is possible to integrate $\Theta $ to obtain a local
action $\Gamma ^{*}$ for the conformal factor $\phi $. 
\begin{eqnarray}
&&\Gamma ^{*}[\phi ]=\frac{1}{840(4\pi )^{3}}\int {\rm d}^{6}x\ \left\{
a^{\prime }\phi \Box ^{3}\phi -(6a^{\prime }-13a_{1}^{\prime
}+7a_{2}^{\prime })(\Box \phi )^{3}+(8a^{\prime }-8a_{2}^{\prime })\Box \phi
(\partial _{\mu }\partial _{\nu }\phi )^{2}\right.  \nonumber \\
&&-(8a+12a^{\prime }-78a_{1}^{\prime }+58a_{2}^{\prime })(\Box \phi
)^{2}(\partial _{\mu }\phi )^{2}+(8a+8a^{\prime }-16a_{2}^{\prime
})(\partial _{\alpha }\phi )^{2}(\partial _{\mu }\partial _{\nu }\phi )^{2} 
\nonumber \\
&&+(8a^{\prime }-8a_{2}^{\prime })(\Box \partial _{\mu }\phi )(\partial
^{\mu }\phi )(\partial _{\alpha }\phi )^{2}-(12a+24a^{\prime
}-156a_{1}^{\prime }+120a_{2}^{\prime })(\partial _{\mu }\phi )^{4}\Box \phi
\nonumber \\
&&\left. -(8a+16a^{\prime }-104a_{1}^{\prime }+80a_{2}^{\prime })(\partial
_{\mu }\phi )^{6}\right\} .  \label{6corr}
\end{eqnarray}
If $a_{i}^{\prime }=a$ the action $\Gamma ^{*}$ reduces to: 
\[
\Gamma ^{*}[\phi ]=\frac{a}{840~(4\pi )^{3}}\int {\rm d}^{6}x(\phi \Box
^{3}\phi ). 
\]
This nice simplification of $\Gamma $, occurring for a specific choice of
the $a^{\prime }$s, is due to the existence of the pondered Euler density 
\cite{at6d} 
\begin{equation}
\sqrt{g}\,\tilde{{\rm G}}_{6}\equiv \sqrt{g}\left[ {\rm G}_{6}+\nabla
_{\alpha }J_{6}^{\alpha }\right] =48~\Box ^{3}\phi  \label{six}
\end{equation}
where 
\begin{eqnarray*}
J_{6}^{\alpha } &=&-\left( \frac{408}{5}-20\zeta \right) R_{\mu }^{\nu
}\nabla _{\nu }R^{\mu \alpha }-\left( \frac{36}{25}+2\zeta \right) R^{\alpha
\mu }\nabla _{\mu }R \\
&&+~\zeta \nabla ^{\alpha }R^{2}+\left( \frac{144}{5}-10\zeta \right) \nabla
^{\alpha }(R_{\mu \nu }R^{\mu \nu })-\frac{24}{5}\nabla ^{\alpha }\Box R
\end{eqnarray*}
and $\zeta $ is arbitrary.

Using the procedure of \cite{234} we can derive from (\ref{6corr}) a set of
equivalent sum rules for $\Delta a$ and the $\Delta a_{j}^{\prime }$s and a
set of vanishing relations. The ``minimal'' sum rule for $\Delta a$ is 
\begin{eqnarray}
\Delta a &=&\frac{7\pi ^{3}}{36}\left( -6\int x^{6}\;\Gamma _{x,0}^{\prime
}\;{\rm d}^{6}x+\int [8(x\cdot y)^{3}-9x^{4}y^{2}]\;\Gamma _{x,y,0}^{\prime
}\;{\rm d}^{6}x{\rm d}^{6}y\right.  \nonumber \\
&&\qquad \qquad \qquad \qquad \qquad \left. +6\int x^{2}(y\cdot
z)^{2}\;\Gamma _{x,y,z,0}^{\prime }\;{\rm d}^{6}x{\rm d}^{6}y{\rm d}%
^{6}z\right) .  \label{minimale}
\end{eqnarray}
In arbitrary even dimensions $2n$ the analogue minimal sum rule will contain
terms up to the ($n+1$)th derivative of $\Gamma ^{\prime }$. The increasing
complexity of the formulas with the spacetime dimension reflects the
increasing complexity of the structure of the action in external gravity.

If we ignore the existence of an embedding in external gravity, the most
general form for $\Gamma ^{*}$, subject only to the constraint that it is
polynomial in $\phi $ and its derivatives, and free of dimensionful
parameters, is 
\begin{eqnarray}
\Gamma ^{*}[\phi ] &=&\int {\rm d}^{6}x\ (b_{1}\phi \Box ^{3}\phi
+b_{2}(\Box \phi )^{3}+b_{3}\Box \phi (\partial _{\mu }\partial _{\nu }\phi
)^{2}+b_{4}(\Box \phi )^{2}(\partial _{\mu }\phi )^{2}+b_{5}(\partial
_{\alpha }\phi )^{2}(\partial _{\mu }\partial _{\nu }\phi )^{2}  \nonumber \\
&&+b_{6}(\Box \partial _{\mu }\phi )(\partial ^{\mu }\phi )(\partial
_{\alpha }\phi )^{2}+b_{7}(\partial _{\mu }\phi )^{4}\Box \phi
+b_{8}(\partial _{\mu }\phi )^{6}).  \label{6cin}
\end{eqnarray}
We can write a set of sum rules for $\Gamma ^{\prime }$ expressing the
difference between the values ofthe $b_{i}$s at the critical points. They
are collected in the appendix. We know, from the embedding in external
gravity, that only 4 $b_{i}$s are independent, so there must be at least
four vanishing relations among the $\Delta b_{i}$s. These can be viewed as
trivial flow invariants. Moreover, we know that another linear combination
of the $\Delta b_{i}$s is flow invariant, namely $\Delta a$. This makes at
least five flow invariants in total, predicted by the embedding in external
gravity. This counting is confirmed by our method based only on flow
invariance.

We study the sum rules in the model (\ref{azionenonunit}). Expressing the
action (\ref{azionenonunit}) by means of $\stackrel{\sim }{\varphi }\equiv
\varphi {\rm e}^{\phi }$, choosing a conformally flat metric and
functionally differentiating with respect to $\phi $ at fixed $\stackrel{%
\sim }{\varphi }$, we arrive at the stress-tensor trace up to terms
proportional to the field equations: 
\begin{eqnarray*}
&&\stackrel{\sim }{\Theta }=-\beta m^{2}{\rm e}^{2\phi }((\partial _{\mu }%
\stackrel{\sim }{\varphi })^{2}+\frac{1}{2}\Box \stackrel{\sim }{\varphi }%
^{2}-\Box (\phi )\stackrel{\sim }{\varphi }^{2}-(\partial _{\mu }\phi )^{2}%
\stackrel{\sim }{\varphi }^{2}-(\partial _{\mu }\phi )(\partial ^{\mu }%
\stackrel{\sim }{\varphi }^{2})) \\
&&-2m^{4}{\rm e}^{4\phi }\stackrel{\sim }{\varphi }^{2}+5\eta m^{2}{\rm e}%
^{2\phi }\Box (\stackrel{\sim }{\varphi }^{2}).
\end{eqnarray*}
The terms proportional to the field equations cancel out in the combinations
(\ref{dera}). We can choose the improvement parameter $\eta $ as we wish,
since the improvement term vanishes at both critical points and flow
invariants are $\eta $-independent, as shown in \cite{impr}. The most
convenient choice for practical calculations is $\eta =\frac{1}{10}\beta $,
in which case the trace operator reduces to 
\[
\stackrel{\sim }{\Theta }=-\beta m^{2}{\rm e}^{2\phi }((\partial _{\mu }%
\stackrel{\sim }{\varphi })^{2}-\Box (\phi )\stackrel{\sim }{\varphi }%
^{2}-(\partial _{\mu }\phi )^{2}\stackrel{\sim }{\varphi }^{2}-(\partial
_{\mu }\phi )(\partial ^{\mu }\stackrel{\sim }{\varphi }^{2}))-2m^{4}{\rm e}%
^{4\phi }\stackrel{\sim }{\varphi }^{2} 
\]

We can evaluate explicitly the sum rules for the $\Delta b_{i}$. For our
purposes it is sufficient to compute $\Delta b_{1,\cdots 6}$ . The results
are given in the appendix. Three combinations $\sum_{i=1}^{6}\lambda
_{i}\Delta b_{i}(r)$ are independent of $r$: 
\begin{eqnarray*}
&&\Delta b_{3}-\Delta b_{6}=0, \\
&&6\Delta b_{2}+4\Delta b_{3}-\Delta b_{4}-\Delta b_{5}=0,
\end{eqnarray*}
and 
\begin{equation}
\Delta a=6720\pi ^{3}\left( 8\Delta b_{1}+6\Delta b_{2}+2\Delta b_{3}-\Delta
b_{4}\right) =-\frac{16}{9}.  \label{16/9}
\end{equation}
In the next section this value is verified computing the trace anomaly
explicitly.

In summary, we have found 3 flow invariants evaluating 6 sum rules. This
implies that the remaining two $\Delta b_{i}$s cannot produce more than five
flow invariants in total, precisely the five flow invariants predicted by
the embedding in external gravity. The remaining two vanishing relations
among the $\Delta b_{i}$s are 
\begin{eqnarray*}
2\Delta b_{7}-3\Delta b_{8} &=&0, \\
2\Delta b_{7}+\Delta b_{3}-4\Delta b_{4}-\Delta b_{5} &=&0.
\end{eqnarray*}
Finally, the quantities $\Delta a^{\prime}$ and $a_{1,2}^{\prime }$ are not
flow invariant, as expected by the ambiguities (\ref{ambi}) associated with
them. They are also independent from one another. Their expressions are 
\begin{eqnarray*}
\Delta a^{\prime } &=&53760\pi ^{3}\Delta b_{1}, \\
\Delta a_{1}^{\prime } &=&\frac{6720}{13}\pi ^{3}\left( 104\Delta
b_{1}+8\Delta b_{2}-7\Delta b_{3}\right) , \\
\Delta a_{2}^{\prime } &=&6720\pi ^{3}\left( 8\Delta b_{1}-\Delta
b_{3}\right) .
\end{eqnarray*}

\subsection{Calculation of $a$ for higher-derivative scalar fields}

We use techniques similar to those of \cite{appollonio}. The induced
generating functional $\Gamma [g_{\mu \nu }]$ for the external metric is
defined by 
\[
{\rm \exp }\left( -\Gamma [g_{\mu \nu }]\right) =\int [{\rm d}\varphi ]~{\rm %
\exp }\left( -S[\varphi ,g_{\mu \nu }]\right) , 
\]
where the action is 
\[
S[\varphi ,g_{\mu \nu }]=\frac{1}{2}\int {\rm d}^{2n}x~\sqrt{g}~\varphi
\Delta _{4}\varphi , 
\]
and $2n=d$ is the spacetime dimension.

We have 
\[
\Gamma [g_{\mu \nu }]=\frac{1}{2}\ln \det \Delta _{4},\qquad \left\langle
\Theta \right\rangle =-\left. \frac{\delta \Gamma [{\rm e}^{2\phi }g_{\mu
\nu }]}{\delta \phi }\right| _{\phi =0}. 
\]
If $\varpi _{k}$ denote the eigenvalues of $\Delta _{4}$ with degeneracies $%
\delta _{k}$, we can write, using the zeta-function regularization 
\[
\Gamma [g_{\mu \nu }]=\frac{1}{2}\sum_{k=0}^{\infty }\delta _{k}\ln \varpi
_{k}=-\frac{1}{2}\left. \frac{{\rm d}}{{\rm d}s}{\rm tr[}\Delta
_{4}^{-s}]\right| _{s=0}=-\frac{1}{2}\left. \frac{{\rm d}}{{\rm d}s}%
\sum_{k=0}^{\infty }\frac{\delta _{k}}{\varpi _{k}^{s}}\right| _{s=0}\equiv -%
\frac{1}{2}\left. \frac{{\rm d}\Upsilon \left( s\right) }{{\rm d}s}\right|
_{s=0}. 
\]
The first sum is only formal, because it diverges. Since $\Delta _{4}$
transforms as $\Delta _{4}\rightarrow e^{-4\phi }\Delta _{4}$ under a
conformal rescaling $g_{\mu \nu }\rightarrow {\rm e}^{2\phi }g_{\mu \nu }$
of the metric, we have 
\begin{equation}
\int {\rm d}^{2n}x~\left\langle \Theta (x)\right\rangle =-\int {\rm d}%
^{2n}x\ \left. \frac{\delta \Gamma [{\rm e}^{2\phi }g_{\mu \nu }]}{\delta
\phi }\right| _{\phi =0}=2\left. \frac{{\rm d}}{{\rm d}s}\sum_{k=0}^{\infty
}s\frac{\delta _{k}}{\varpi _{k}^{s}}\right| _{s=0}=2\Upsilon \left(
0\right) .  \label{uno}
\end{equation}

We now work an a sphere, where the formulas simplify considerably. Using (%
\ref{uno}) and (\ref{trace}) and keeping in mind that the Euler
characteristic of the sphere is equal to 2, we obtain

\[
{a_{d}}=(-1)^{n+1}\frac{\,(2n+1)!}{{n}!\left( n-1\right) !}\Upsilon \left(
0\right) . 
\]
Now we compute $\Upsilon \left( 0\right) $ on the sphere. We first observe
that on Einstein spaces, which satisfy $R_{\mu \nu }=\Lambda g_{\mu \nu }$,
the operator $\Delta _{4}$ factorizes into 
\[
\Delta _{4}=\left( \Box -\Lambda \frac{d\left( d-2\right) }{4\left(
d-1\right) }\right) \left( \Box -\Lambda \frac{\left( d-4\right) \left(
d+2\right) }{4\left( d-1\right) }\right) =\Box _{s}\left( \Box _{s}+\frac{%
2\Lambda }{d-1}\right) , 
\]
where 
\[
-\Box _{s}=-\Box +\Lambda \frac{d\left( d-2\right) }{4\left( d-1\right) } 
\]
is the kinetic operator of a unitary scalar field. Since $-\Box _{s}$ has a
complete spectrum of eigenfunctions with positive eigenvalues $\lambda
_{k}=\Lambda (k+d/2)(k+d/2-1)/\left( d-1\right) $, $\Delta _{4}$ has the
same spectrum of egenfunctions and its eigenvalues are $\varpi _{k}=\lambda
_{k}\left( \lambda _{k}-2/(d-1)\right) $, with the same degeneracy $\delta
_{k}$ as the $\lambda _{k}$s. We have 
\begin{eqnarray*}
\varpi _{k} &=&\frac{\Lambda ^{2}}{\left( d-1\right) ^{2}}\left( k+\frac{d}{2%
}-2\right) \left( k+\frac{d}{2}-1\right) \left( k+\frac{d}{2}\right) \left(
k+\frac{d}{2}+1\right) ,\qquad \\
\delta _{k} &=&(2k+d-1)\frac{\left( k+d-2\right) !}{k!\left( d-1\right) !}.
\end{eqnarray*}
We can take $\Lambda =d-1$ to simplify some formulas. We obtain the sum 
\[
\Upsilon \left( s\right) =\sum_{k=0}^{\infty }\frac{\delta _{k}}{\varpi
_{k}^{s}}=\frac{1}{\left( d-1\right) !}\sum_{k=0}^{\infty }\frac{%
(2k+d-1)~\left( k+d-2\right) !}{k!\left( k+n-2\right) ^{s}\left(
k+n-1\right) ^{s}\left( k+n\right) ^{s}\left( k+n+1\right) ^{s}}. 
\]
The $s\rightarrow 0$ limit of this expression can be evaluated as follows.
We write

\[
1+\frac{1}{\left( d-1\right) !}\sum_{k=1}^{\infty }\frac{(2k+d-1)~\left(
k+d-2\right) !}{k!~k^{4s}}\left( 1-s\sum_{i=1}^{4}\ln \left( 1+\frac{n+2-i}{k%
}\right) \right) . 
\]
Higher powers in $s$ do not contribute in the limit, since all terms can be
expressed by means of the Riemann zeta function, which has at worst a simple
pole for $s\rightarrow 0$. We now expand the logarithms in series. Since the
logarithms are multiplied by $s$, we just need to capture the poles of the
zeta functions, which come from sums of the form $\sum_{k=1}^{\infty }1/k$.
For this reason, it is sufficient to truncate the series to $2n$: 
\[
1+\frac{1}{\left( d-1\right) !}\sum_{k=1}^{\infty }\frac{(2k+d-1)~\left(
k+d-2\right) !}{k!~k^{4s}}\left( 1-s\sum_{i=1}^{4}\sum_{j=1}^{2n}\frac{%
(-1)^{j+1}}{j}\left( \frac{n+2-i}{k}\right) ^{j}\right) . 
\]

The $s\rightarrow 0$ limit of this sum can be evaluated using the analytic
continuation of the zeta function in $s=0$. A few results are: 
\[
\begin{tabular}{|c|c|c|c|c|c|c|c|}
\hline
$d$ & 4 & 6 & 8 & 10 & 12 & 14 & 16 \\ 
$a_{d}$ & $-\frac{28}{3}$ & $-\frac{16}{9}$ & $-\frac{52}{45}$ & $-\frac{124%
}{135}$ & $-\frac{56302}{70875}$ & $-\frac{30544}{42525}$ & $-\frac{2977778}{%
4465125}$%
\end{tabular}
\]
Note that the values are all negative. The value $a_{6}=-16/9$ agrees with
the result (\ref{16/9}) coming from the evaluation of the sum rule (\ref
{minimale}), while $a_{4}=-28/3$ agrees with (\ref{auto}).

\section{Search for flow invariants in odd dimensions}

In this section we study the sum rules (\ref{summa}) in three- and
one-dimensional flows. In three dimensions we study again the model (\ref
{azionenonunit}) and find a non-trivial vanishing relation. In one dimension
we study some simple Calogero-type models. In either case, we do not find
nonvanishing flow invariants.

\subsection{Three dimensions}

The stress-tensor trace is $\widetilde{\Theta }=-\widetilde{\delta }S/%
\widetilde{\delta }\phi $ with the derivative taken at constant $\tilde{%
\varphi}=\varphi {\rm e}^{-\phi /2}$ and reads 
\begin{eqnarray}
\widetilde{\Theta } &=&-\beta m^{2}\left( (\partial _{\mu }\tilde{\varphi}%
)^{2}-\frac{1}{4}\Box \tilde{\varphi}^{2}-\frac{1}{4}(\partial _{\mu }\phi
)^{2}\tilde{\varphi}^{2}-\frac{1}{4}\Box \phi \tilde{\varphi}^{2}-\frac{1}{4}%
\partial _{\mu }\phi \partial ^{\mu }\tilde{\varphi}^{2}\right) {\rm e}%
^{2\phi }+  \label{tetani} \\
&&-\eta m^{2}\left( \Box \tilde{\varphi}^{2}+3(\partial _{\mu }\phi )^{2}%
\tilde{\varphi}^{2}+3\Box \phi \tilde{\varphi}^{2}+3\partial _{\mu }\phi
\partial ^{\mu }\tilde{\varphi}^{2}\right) {\rm e}^{2\phi }-2m^{4}\tilde{%
\varphi}^{2}{\rm e}^{4\phi }.  \nonumber
\end{eqnarray}
The field $\tilde{\varphi}$ has dimension $-1/2$ in units of mass and $\phi $
is dimensionless. The most convenient choice for the improvement parameter
is $\eta =1/12$, in which case all derivatives of the conformal factor $\phi 
$ disappear: 
\begin{equation}
\widetilde{\Theta }=-\beta m^{2}((\partial _{\mu }\tilde{\varphi})^{2}-\frac{%
1}{6}\Box \tilde{\varphi}^{2}){\rm e}^{2\phi }-2m^{4}\tilde{\varphi}^{2}{\rm %
e}^{4\phi }.  \label{teta3}
\end{equation}
We have, in flat space 
\[
\frac{\delta \widetilde{\Theta }(x)}{\delta \phi (y)}=\delta (x-y)\left[ 2%
\widetilde{\Theta }(x)-4m^{4}\tilde{\varphi}^{2}(x)\right] ,\qquad \frac{%
\delta ^{2}\widetilde{\Theta }(x)}{\delta \phi (y)\delta \phi (z)}=\delta
(x-y)\delta (x-z)\left[ 4\widetilde{\Theta }(x)-24m^{4}\tilde{\varphi}%
^{2}(x)\right] . 
\]

The functions $P_{i}\left( x_{1},\cdots ,x_{i}\right) $ of the ansatz (\ref
{summa}) cannot be polynomials in odd dimensions. The most general sum rule
contains infinitely many terms. The best we can do is to explore a subclass
of sum rules in which the $P_{i}$s are particularly simple, such that (\ref
{summa}) contains finitely many terms. For example, we can take the $P_{i}$s
which are the product of a polynomial of degree 2 and the modulus of a
distance between the points $x_{1},\cdots ,x_{i}$ and $0$. We recall that
the equivalence relations \cite{234} 
\begin{eqnarray*}
P_{i}(x_{1},\cdots ,x_{i}) &\sim &P_{i}(x_{1}-x_{k},\cdots
,x_{k-1}-x_{k},-x_{k},x_{k+1}-x_{k},\cdots ,x_{i}-x_{k})\qquad \forall
k=1,\cdots ,i, \\
P_{i}(x_{1},\cdots ,x_{i}) &\sim &0\qquad \text{if }P_{i}\text{ is
independent of any of the }x_{k}.
\end{eqnarray*}
reduce the set of independent functions. Our subclass contains only three
terms: for $i=1$ we have only $P_{1}\left( x\right) =|x|^{3}$; for $i=2$ we
have only $P_{2}\left( x,y\right) =|x|\ y^{2}$, since $P_{2}\left(
x,y\right) =|x|^{3}\sim 0,$ $|x|\ x\cdot y\sim 0,$ $|x-y|\ x^{2}\sim |x-y|\
x\cdot y\sim |x|\ y^{2}$; for $i=3$ we have only $P_{3}\left( x,y,z\right)
=|x|\ y\cdot z,$ since $|x-y|\ z^{2}\sim 2|x-y|\ x\cdot z\sim -2|x|\ y\cdot
z $; for $i>3$ there is no nontrivial fuction $P_{i}$ in our subclass. We
have to study the expression 
\begin{equation}
\Delta {\cal A}=-\lambda _{1}\int |x|^{3}\ \Gamma _{x,0}^{\prime }\ {\rm d}%
^{3}x-\lambda _{2}\int |x|\ y^{2}\ \Gamma _{x,y,0}^{\prime }\ {\rm d}^{3}x%
{\rm d}^{3}y-\lambda _{3}\int |x|\ y\cdot z\ \Gamma _{x,y,z,0}^{\prime }\ 
{\rm d}^{3}x{\rm d}^{3}y{\rm d}^{3}z.  \label{sum3}
\end{equation}

With the notation 
\begin{eqnarray*}
\pi {\rm T}_{2} &=&\int |x|^{3}\left\langle \widetilde{\Theta }(x)\widetilde{%
\Theta }(0)\right\rangle \ {\rm d}^{3}x \\
\pi {\rm TF} &=&m^{4}\int |x|^{3}\left\langle \widetilde{\Theta }(x)\tilde{%
\varphi}^{2}(0)\right\rangle \ {\rm d}^{3}x \\
\pi {\rm T}_{3} &=&\int |x|y^{2}\left\langle \widetilde{\Theta }(x)%
\widetilde{\Theta }(y)\widetilde{\Theta }(0)\right\rangle \ {\rm d}^{3}x{\rm %
d}^{3}y \\
\pi {\rm T}_{4} &=&\int |x|\left( y\cdot z\right) \left\langle \widetilde{%
\Theta }(x)\widetilde{\Theta }(y)\widetilde{\Theta }(z)\widetilde{\Theta }%
(0)\right\rangle \ {\rm d}^{3}x{\rm d}^{3}y{\rm d}^{3}z \\
\pi {\rm T}_{2}{\rm F}_{1} &=&m^{4}\int |x|y^{2}\left\langle \widetilde{%
\Theta }(x)\widetilde{\Theta }(y)\tilde{\varphi}^{2}(0)\right\rangle \ {\rm d%
}^{3}x{\rm d}^{3}y \\
\pi {\rm T}_{2}{\rm F}_{2} &=&m^{4}\int |x|\left( x\cdot y\right)
\left\langle \widetilde{\Theta }(x)\widetilde{\Theta }(y)\tilde{\varphi}%
^{2}(0)\right\rangle \ {\rm d}^{3}x{\rm d}^{3}y
\end{eqnarray*}
(in every correlator $\widetilde{\Theta }$ is meant in flat space, i.e. at $%
\phi =0$), we find 
\begin{eqnarray*}
{\rm I}_{1} &=&-\int |x|^{3}\ \Gamma _{x,0}^{\prime }\ {\rm d}^{3}x=\pi {\rm %
T}_{2}{\rm ,} \\
{\rm I}_{2} &=&-\int |x|\ y^{2}\ \Gamma _{x,y,0}^{\prime }\ {\rm d}^{3}x{\rm %
d}^{3}y=\pi \left( {\rm T}_{3}+2{\rm T}_{2}-4{\rm TF}\right) {\rm ,} \\
{\rm I}_{3} &=&-\int |x|\ y\cdot z\ \Gamma _{x,y,z,0}^{\prime }\ {\rm d}^{3}x%
{\rm d}^{3}y{\rm d}^{3}z=\pi \left( {\rm T}_{4}+2{\rm T}_{3}-4{\rm T}_{2}%
{\rm F}_{1}-8{\rm T}_{2}{\rm F}_{2}-4{\rm T}_{2}-8{\rm TF}\right) .
\end{eqnarray*}
To write the last formula, we have used a kinematic vanishing relation of
the type \cite{234}, namely 
\[
\int {\rm d}^{2n}x\,\prod_{i=1}^{k-1}{\rm d}^{2n}x_{i}\,F(x_{1},\cdots
,x_{k-1})\,\Gamma _{xx_{1}\cdots x_{k-1}0}^{\prime }=0. 
\]
This identity holds for every homogeneous function $F(x_{1},\cdots ,x_{k-1})$
of degree $d=2n$. In particular, we have 
\[
\int |x|\left( x\cdot y\right) \left\langle \widetilde{\Theta }(x)\widetilde{%
\Theta }(y)\widetilde{\Theta }(0)\right\rangle \ {\rm d}^{3}x{\rm d}%
^{3}y=-2\pi {\rm T}_{2}+4\pi {\rm TF} 
\]
To study the $r$-dependence of $\Delta {\cal A}(r)$, we have computed the
values of I$_{1,2,3}$ and their second, forth and sixth derivatives at $r=1$%
(the odd derivatives are related to the even derivatives in a simple way).
We find 
\[
\begin{tabular}{|c|c|c|c|c|c|c|c|c|c|}
& T$_{2}$ & TF & T$_{3}$ & T$_{2}$F$_{1}$ & T$_{2}$F$_{2}$ & T$_{4}$ & I$%
_{1}/\pi $ & I$_{2}/\pi $ & I$_{3}/\pi $ \\ 
$r=1$ & $\frac{130}{3}$ & $-12$ & $-\frac{1858}{9}$ & $\frac{3208}{45}$ & $%
48 $ & $\frac{50557}{45}$ & $\frac{130}{3}$ & $-\frac{646}{9}$ & $-\frac{323%
}{9} $ \\ 
${\rm d}^{2}/{\rm d}r^{2}$ at $r=1$ & $\frac{16}{3}$ & $\frac{1}{2}$ & $-%
\frac{26}{3}$ & $-\frac{4483}{945}$ & $-2$ & $\frac{7268}{945}$ & $\frac{16}{%
3}$ & $0$ & $0$ \\ 
${\rm d}^{4}/{\rm d}r^{4}$ at $r=1$ & $\frac{43}{9}$ & $\frac{5}{16}$ & $-%
\frac{865}{108}$ & $-\frac{100399}{30240}$ & $-\frac{5}{4}$ & $\frac{109601}{%
7560}$ & $\frac{43}{9}$ & $\frac{8}{27}$ & $\frac{4}{27}$ \\ 
${\rm d}^{6}/{\rm d}r^{6}$ at $r=1$ & $\frac{179}{48}$ & $0$ & $-\frac{953}{%
144}$ & $-\frac{610121}{665280}$ & $0$ & $\frac{8284253}{332640}$ & $\frac{%
179}{48}$ & $\frac{121}{144}$ & $\frac{121}{288}$%
\end{tabular}
\]

Analysing the data of this table, we do not find non-trivial flow
invariants, but we do find a non-trivial vanishing relation, which reads 
\begin{equation}
\int |x|\ y^{2}\ \Gamma _{x,y,0}^{\prime }\ {\rm d}^{3}x{\rm d}^{3}y-2\int
|x|\ y\cdot z\ \Gamma _{x,y,z,0}^{\prime }\ {\rm d}^{3}x{\rm d}^{3}y{\rm d}%
^{3}z=0.  \label{van}
\end{equation}
Presumably, it is possible to prove this relation in full generality
studying the embedding in external gravity. This does not seem to be
straightforward, however, since the $\phi $-dependence is trivial at the
critical points and non-local at intermediate energies.

\subsubsection{Improvement}

To perform the calculations, we have chosen a convenient value $\eta =1/12$
of the improvement parameter in (\ref{tetani}). Due to the non-locality of
the functions $P_{i}(x_{1},\cdots ,x_{i})$, the other values of $\eta $
produce undesirable divergent terms of the form 
\begin{equation}
\int {\rm d}^{3}x\frac{\delta (x)}{|x|}G(|x|),  \label{diva}
\end{equation}
in the integrals I$_{2}$ and I$_{3}$. Here $G(|x|)$ denotes some regular
function. It was shown in \cite{impr} that a flow invariant is insensitive
to the value of the improvement parameter $\eta $. So, we expect that the
divergences (\ref{diva}) cancel out in the combination (\ref{van}). This
happens if $G(0)=0$. We are going to check that this is true inspecting the
terms of the form (\ref{diva}) in (\ref{van}) to the first order in $%
1-12\eta $.

Observe that because of the $\delta $-function in (\ref{diva}), the
divergences (\ref{diva}) appear only in the integrals of correlators
containing insertions of $\delta \widetilde{\Theta }(x)/\delta \phi
(y)|_{\phi =0}$ and $\delta ^{2}\widetilde{\Theta }(x)/\delta \phi (y)\delta
\phi (z)|_{\phi =0}$. They do not appear in the integrals of correlators
containing only insertions of $\widetilde{\Theta }$. In particular, no
divergence (\ref{diva}) appears in I$_{1}$. Performing the differentiations,
we find that, to the first order in $1-12\eta $, $G(0)$ is equal to $%
1-12\eta $ times 
\begin{equation}
2\int {\rm d}^{3}y{\rm d}^{3}z\ y\cdot z\ \langle \widetilde{\Theta }(y)\ 
\widetilde{\Theta }(z)\ {\normalsize \tilde{\varphi}^{2}}(0)\rangle
+m^{2}\int {\rm d}^{3}y\ y^{2}\ \left\langle \left( 3\widetilde{\Theta }%
(y)-8m^{2}{\normalsize \tilde{\varphi}^{2}}(y)\right) \ {\normalsize \tilde{%
\varphi}^{2}}(0)\right\rangle  \label{comb}
\end{equation}

Now, these integrals contain only local polynomials and can be computed
easily for arbitrary $r$. We find 
\begin{eqnarray*}
\int {\rm d}^{3}y{\rm d}^{3}z\ y\cdot z\ \langle \widetilde{\Theta }(y)\ 
\widetilde{\Theta }(z)\ {\normalsize \tilde{\varphi}^{2}}(0)\rangle &=&\frac{%
\sqrt{r}\left( 23r^{4}+100r^{3}+186r^{2}+100r+23\right) }{16(r+1)^{5}}, \\
m^{2}\int {\rm d}^{3}y\ y^{2}\ \langle \widetilde{\Theta }(y)\ {\normalsize 
\tilde{\varphi}^{2}}(0)\rangle &=&-\frac{5\sqrt{r}}{8(r+1)}, \\
m^{4}\int {\rm d}^{3}y\ y^{2}\ \langle {\normalsize \tilde{\varphi}^{2}}(y)\ 
{\normalsize \tilde{\varphi}^{2}}(0)\rangle &=&\frac{\sqrt{r}\left(
r^{4}+5r^{3}+12r^{2}+5r+1\right) }{8(r+1)^{5}}.
\end{eqnarray*}
The combination (\ref{comb}) is the unique vanishing linear combination of
the three integrals just computed.

This result is further evidence that the vanishing of (\ref{van}) is not a
coincidence, but must be a consequence of the embedding in external gravity.
Moreover, the convergence criterion $G(0)=0$ can be used to fix the relation
(\ref{van}) with much less effort than the inspection of flow invariance. We
expect that, with the methods developed in this section, many more
non-trivial relations can be found.

\subsection{One dimension}

In one dimension the stress tensor coincides with its trace. The stress
tensor two-point function does not define a central charge $c$. We therefore
look for a central charge of type $a$. The simplest candidate sum rule for
the difference $\Delta {\cal A}$ between the critical values of a central
charge of type $a$ reads

\[
\Delta {\cal A}=\int_{-\infty }^{+\infty }{\rm d}t\ |t|\ \left\langle
0\left| T\left( {\rm {:}}\Theta (t){\rm {:}\ {:}}\Theta (0){\rm {:}}\right)
\right| 0\right\rangle =2\int_{0}^{\infty }{\rm d}t\ t\ \left\langle 0\left| 
{\rm {:}}\Theta (t){\rm {:}\ {:}}\Theta (0){\rm {:}}\right| 0\right\rangle . 
\]
Here ${\rm {:}}\Theta (t){\rm {:}}=\Theta (t)-\left\langle 0\left| \Theta
(0)\right| 0\right\rangle $ is the normal product. Inserting a complete set
of orthonormalized states $\left| \left. n\right\rangle \right. $, we obtain 
\begin{equation}
\Delta {\cal A}=2\int_{0}^{\infty }{\rm d}t\ t\ \sum_{n}\left\langle 0\left| 
{\rm {:}}\Theta (t){\rm {:}}\right| n\right\rangle \left\langle n\left| {\rm 
{:}}\Theta (0){\rm {:}}\right| 0\right\rangle =2\sum_{n>0}\frac{\left|
\left\langle 0\left| \Theta \right| n\right\rangle \right| ^{2}}{\left(
E_{n}-E_{0}\right) ^{2}}.  \label{sumo}
\end{equation}

We want to test the flow invariance of this expression in a family of flows
interpolating between the same fixed points, or between two families of
continuously connected fixed points. In one dimension, the classically
conformal potential is 
\[
V(q)=\frac{1}{q^{2}}. 
\]
We consider the model 
\begin{equation}
{\cal L}=\frac{1}{2}\sum_{i=1}^{2}\left( \dot{q}_{i}^{2}+m^{2}q_{i}^{2}%
\right) +\frac{g^{2}}{2q^{2}},  \label{m1}
\end{equation}
where $q^{2}=q_{1}^{2}+q_{2}^{2}$ and $g$ is positive. We prefer to study
this model, instead of the ``two-body Calogero model'' 
\begin{equation}
{\cal L}=\frac{1}{2}\left( \dot{q}^{2}+m^{2}q^{2}\right) +\frac{g^{2}}{2q^{2}%
},  \label{m2}
\end{equation}
since (\ref{m1}) reduces to a harmonic oscillator for $g=0$, while (\ref{m2}%
) does not. We use the Euclidean notation. We have a potential 
\[
V=\frac{1}{2}\left( m^{2}q^{2}+\frac{g^{2}}{q^{2}}\right) , 
\]
with minima at 
\[
\overline{q}=\pm \sqrt{\frac{g}{m}},\qquad V\left( \overline{q}\right) =mg. 
\]
The stress tensor is equal to its trace: 
\[
\Theta =-m^{2}q^{2}. 
\]
The eigenfunctions and eigenvalues are 
\begin{equation}
\psi _{n,l}\left( \rho ,\theta \right) =\sqrt{\frac{m^{\nu +1}n!}{\pi \left(
n+\nu \right) !}}\ {\rm e}^{il\theta -m\rho ^{2}/2}\rho ^{\nu }L_{n}^{\nu
}\left( m\rho ^{2}\right) ,\qquad E_{n,l}=m\left( 2n+1+\nu \right) ,
\label{fonda}
\end{equation}
with $\nu =\sqrt{g^{2}+l^{2}}$. Here $L_{n}^{\nu }$ is the associated
Laguerre polynomial. When $g\rightarrow 0$ we recover the two-dimensional
harmonic oscillator.

The UV\ fixed points of the model (\ref{m1}) are the one-parameter family of
conformal field theories 
\[
{\cal L}=\frac{1}{2}\sum_{i=1}^{2}\dot{q}_{i}^{2}+\frac{g^{2}}{2q^{2}}. 
\]
If the central charge ${\cal A}_{{\rm UV}}$ satisfied the marginality
property 3) of the introduction, then it would be $g$-independent.

The IR\ fixed point of (\ref{m1}) is the empty theory, since the potential
is everywhere infinite for $m\rightarrow \infty $. We therefore expect $%
{\cal A}_{{\rm IR}}=0$ for every candidate central charge ${\cal A}$. We
conclude that marginality and flow invariance require that $\Delta {\cal A}$
does not depend on $g$. We eant to check if our candidate $\Delta {\cal A}$
of (\ref{sumo}) is $g$-independent.

This does not turn out to be true. Indeed, using some standard properties of
the associated Laguerre polynomials, we find 
\begin{equation}
\left\langle 0\left| q^{2}\right| n,l\right\rangle =\frac{\left( g+1\right) !%
}{m}\sqrt{\frac{n!}{g!\left( n+g\right) !}}\delta _{l0}\left( \delta
_{n0}-\delta _{n1}\right)  \label{media}
\end{equation}
and 
\begin{equation}
\Delta {\cal A}=\frac{1}{2}(1+g).  \label{po}
\end{equation}
We can try with more general sum rules. Straightforward manipulations allow
us to derive the generalization of formula (\ref{media}), which reads

\[
\left\langle n,l\left| \Theta \right| n^{\prime },l^{\prime }\right\rangle
=-m\delta _{ll^{\prime }}\left[ \delta _{nn^{\prime }}(2n+\nu +1)-\sqrt{%
(n+1)(n+\nu +1)}\delta _{n+1,n^{\prime }}-\sqrt{n(n+\nu )}\delta
_{n-1,n^{\prime }}\right] . 
\]
Using this result, we can compute

\[
\sum\Sb n,n^{\prime },n^{\prime \prime }>0  \\ n\neq n^{\prime }\neq
n^{\prime \prime }  \endSb \frac{\left\langle 0\left| \Theta \right|
n,l\right\rangle }{E_{0}-E_{n}}\frac{\left\langle n,l\left| \Theta \right|
n^{\prime },l^{\prime }\right\rangle }{E_{n}-E_{n^{\prime }}}\frac{%
\left\langle n^{\prime },l^{\prime }\left| \Theta \right| n^{\prime \prime
},l^{\prime \prime }\right\rangle }{E_{n^{\prime }}-E_{n^{\prime \prime }}}%
\frac{\left\langle n^{\prime \prime },l^{\prime \prime }\left| \Theta
\right| 0\right\rangle }{E_{n^{\prime \prime }}-E_{0}}=-\frac{1}{16}\left(
5+8g+3g^{2}\right) . 
\]
A natural temptative sum rule containing infinitely many terms reads 
\[
\Delta {\cal A}=2\sum_{n>0}\frac{\left| \left\langle 0\left| \Theta \right|
n\right\rangle \right| ^{2}}{\left( E_{n}-E_{0}\right) ^{2}}+d_{1}\sum\Sb %
n,n^{\prime },n^{\prime \prime }>0  \\ n\neq n^{\prime }\neq n^{\prime
\prime }  \endSb \frac{\left\langle 0\left| \Theta \right| n,l\right\rangle 
}{E_{0}-E_{n}}\frac{\left\langle n,l\left| \Theta \right| n^{\prime
},l^{\prime }\right\rangle }{E_{n}-E_{n^{\prime }}}\frac{\left\langle
n^{\prime },l^{\prime }\left| \Theta \right| n^{\prime \prime },l^{\prime
\prime }\right\rangle }{E_{n^{\prime }}-E_{n^{\prime \prime }}}\frac{%
\left\langle n^{\prime \prime },l^{\prime \prime }\left| \Theta \right|
0\right\rangle }{E_{n^{\prime \prime }}-E_{0}}+\cdots , 
\]
where the constants $d_{i}$ should be fixed imposing that $\Delta {\cal A}$
be $g$-independent. However, the infinite system of equations cannot be
solved recursively. This is evident from the expressions (\ref{uno}) and (%
\ref{due}) of the first two contributions. Our procedure does not allow us
to fix a sum rule with infinitely many terms using only flow invariance.

\section{Conclusions}

Several properties of the trace anomalies in external gravity appear to lie
beyond the gravitational embedding and need a more general treatment to be
fully understood. In this paper we have made some steps in this direction.
In even dimensions we have been able to recover the known results without
relying on the gravitational embedding. In odd dimensions the situation is
complicated by the fact that if a non-trivial flow invariant of type $a$
exists, the sum rule (\ref{summa}) probably contains infinitely many terms.
The procedure of this paper does not allow us to explore sum rules
containing infinitely many terms. Within a class of sum rules containing
finitely many terms we find the nontrivial vanishing relation (\ref{van}) in
three dimensions, but we do not find a non-vanishing flow invariant of type $%
a$. Our results are an indication that such a quantity might not exist at
all in odd dimensions.

We mention some directions for future investigations. The study of the
irreversibility of the RG flow in even dimensions \cite{proc,athm,at6d,cea}
points out the peculiarity of classically conformal theories. This class of
theories might be peculiar in odd dimensions also. The simplest
odd-dimensional domain for this study is three-dimensional quantum field
theory, in particular the flows constructed in \cite{largeN}. Indeed, in one
dimension classically conformal theories have trivial flows, because they
are conformal also at the quantum level. Finally, the properties of the
central charge $c$, defined using the stress-tensor two-point function, need
to be studied apart. Progresses in these directions will be reported soon.

\vskip 1truecm

{\bf Appendix. Sum rules in six dimensions}

\vskip .5truecm

Here we collect some lengthy formulas used in section 2.2. The sum rules for
the $b_{i}$s of formula (\ref{6cin}) are 
\begin{eqnarray*}
\int x^{6}\;\Gamma _{x,0}^{\prime }\;{\rm d}^{6}x &=&-46080\Delta b_{1}, \\
\int (x\cdot y)^{3}\;\Gamma _{x,y,0}^{\prime }\;{\rm d}^{6}x{\rm d}^{6}y
&=&-6912\Delta b_{2}-12672\Delta b_{3}, \\
\int x^{4}y^{2}\;\Gamma _{x,y,0}^{\prime }\;{\rm d}^{6}x{\rm d}^{6}y
&=&-27648\Delta b_{2}-12288\Delta b_{3}, \\
\int x^{2}(x\cdot y)(x\cdot z)\;\Gamma _{x,y,z,0}^{\prime }\;{\rm d}^{6}x%
{\rm d}^{6}y{\rm d}^{6}z &=&-1536\Delta b_{4}-1536\Delta b_{5}+8448\Delta
b_{6}, \\
\int x^{4}(y\cdot z)\;\Gamma _{x,y,z,0}^{\prime }\;{\rm d}^{6}x{\rm d}^{6}y%
{\rm d}^{6}z &=&-9216\Delta b_{4}-9216\Delta b_{5}+12288\Delta b_{6}, \\
\int x^{2}(y\cdot z)^{2}\;\Gamma _{x,y,z,0}^{\prime }\;{\rm d}^{6}x{\rm d}%
^{6}y{\rm d}^{6}z &=&-5376\Delta b_{4}+384\Delta b_{5}+8448\Delta b_{6}, \\
\int x^{2}y^{2}(z\cdot k)\;\Gamma _{x,y,z,k,0}^{\prime }\;{\rm d}^{6}x{\rm d}%
^{6}y{\rm d}^{6}z{\rm d}^{6}k &=&15360\Delta b_{7}, \\
\int x^{2}(y\cdot t)(z\cdot k)\;\Gamma _{x,y,z,k,t,0}^{\prime }\;{\rm d}^{6}x%
{\rm d}^{6}y{\rm d}^{6}z{\rm d}^{6}k{\rm d}^{6}t &=&-46080\Delta b_{8}.
\end{eqnarray*}
Their evaluation in the model (\ref{azionenonunit}) gives 
\begin{eqnarray*}
(4\pi )^{3}\Delta b_{1} &=&-\frac{1}{5040(r^{2}-1)^{5}}%
(149r^{10}+25r^{8}-1660{r^{6}}+1660{r^{4}}-25{r^{2}}-149 \\
&&-420(r^{10}-r^{8}-2{r^{6}}-2{r^{4}}-r^{2}+1)\log (r)), \\
(4\pi )^{3}\Delta b_{2} &=&\frac{1}{18900{{({r^{2}}-1)}^{5}}}(3649{r^{10}}%
+9475{r^{8}}-31550{r^{6}}+31550{r^{4}}-9475{r^{2}}-3649 \\
&&-1890(5{r^{10}}-{r^{8}}+4{r^{6}}+4{r^{4}}-{r^{2}}+5)\log (r)), \\
(4\pi )^{3}\Delta b_{3} &=&-\frac{1}{1575{{({r^{2}}-1)}^{5}}}(167{r^{10}}+775%
{r^{8}}-4000{r^{6}}+4000{r^{4}}-775{r^{2}}-167 \\
&&-210(5{r^{10}}-7{r^{8}}-2{r^{6}}-2{r^{4}}-7{r^{2}}+5)\log (r)), \\
(4\pi )^{3}\Delta b_{4} &=&\frac{1}{4725{{({r^{2}}-1)}^{5}}}(3434{r^{10}}%
+8975{r^{8}}-10075{r^{6}}+10075{r^{4}}-8975{r^{2}}-3434 \\
&&-315(15{r^{10}}+29{r^{8}}+64{r^{6}}+64{r^{4}}+29{r^{2}}+15)\log (r)), \\
(4\pi )^{3}\Delta b_{5} &=&\frac{1}{9450{{({r^{2}}-1)}^{5}}}(71{r^{10}}-8125{%
r^{8}}+21500{r^{6}}-21500{r^{4}}+8125{r^{2}}-71 \\
&&+210(5{r^{10}}-9{r^{8}}+6{r^{6}}+6{r^{4}}-9{r^{2}}+5)\log (r)), \\
(4\pi )^{3}\Delta b_{6} &=&-\frac{1}{1575{{({r^{2}}-1)}^{5}}}(167{r^{10}}+775%
{r^{8}}-4000{r^{6}}+4000{r^{4}}-775{r^{2}}-167 \\
&&-210(5{r^{10}}-7{r^{8}}-2{r^{6}}-2{r^{4}}-7{r^{2}}+5)\log (r)).
\end{eqnarray*}

\vskip 1truecm

{\bf Acknowledgements}

\vskip .5truecm

D.A. and G.F. would like to thank the Aspen Center for Physics and the
Center for Theoretical Physics of MIT, respectively, for warm hospitality
during the late stage of this work.

\end{document}